\documentclass[conference]{IEEEtran}
\IEEEoverridecommandlockouts
\usepackage{cite}
\usepackage{amsmath,amssymb,amsfonts}
\usepackage{algorithmic}
\usepackage{graphicx}
\usepackage{textcomp}
\usepackage{xcolor}
\usepackage{booktabs}
\usepackage{listings}
\usepackage{tabularx}
\usepackage{multirow}
\usepackage[table]{xcolor}
\usepackage{array}
\usepackage{xspace}
\usepackage{enumitem}
\usepackage{mfirstuc}
\usepackage{comment}
\usepackage{makecell}
\usepackage[most]{tcolorbox}

\definecolor{codebg}{RGB}{248,248,248}
\definecolor{codeframe}{RGB}{210,210,210}
\definecolor{codekeyword}{RGB}{0,92,175}
\definecolor{codestring}{RGB}{163,21,21}
\definecolor{codecomment}{RGB}{0,128,0}

\lstdefinestyle{javacode}{
    language=Java,
    backgroundcolor=\color{codebg},
    frame=single,
    rulecolor=\color{codeframe},
    framerule=0.4pt,
    framesep=4pt,
    xleftmargin=1.6em,
    xrightmargin=2pt,
    basicstyle=\ttfamily\scriptsize,
    keywordstyle=\color{codekeyword}\bfseries,
    stringstyle=\color{codestring},
    commentstyle=\color{codecomment}\itshape,
    numbers=left,
    numberstyle=\tiny\color{gray!60},
    stepnumber=1,
    numbersep=7pt,
    breaklines=true,
    breakatwhitespace=true,
    showstringspaces=false,
    tabsize=2,
    columns=fullflexible,
    keepspaces=true,
    captionpos=b
}
\def\BibTeX{{\rm B\kern-.05em{\sc i\kern-.025em b}\kern-.08em
    T\kern-.1667em\lower.7ex\hbox{E}\kern-.125emX}}

\lstdefinelanguage{CodeQL}{
  morekeywords={
    import, from, where, select, predicate, exists, or, and, not,
    class, extends, instanceof, result, this, string, int, boolean
  },
  sensitive=true,
  morecomment=[l]{//},
  morecomment=[s]{/*}{*/},
  morestring=[b]",
}

\lstdefinestyle{codeqlstyle}{
  language=CodeQL,
  basicstyle=\ttfamily\scriptsize,
  keywordstyle=\color{codekeyword}\bfseries,
  commentstyle=\color{codecomment}\itshape,
  stringstyle=\color{codestring},
  backgroundcolor=\color{codebg},
  rulecolor=\color{codeframe},
  columns=fullflexible,
  keepspaces=true,
  showstringspaces=false,
  breaklines=true,
  breakatwhitespace=true,
  frame=single,
  xleftmargin=0.5em,
  xrightmargin=0.5em,
  aboveskip=4pt,
  belowskip=4pt,
  captionpos=b
}

\newcolumntype{?}{!{\vrule width 1pt}}
\newcommand{\mypara}[1]{\noindent\textbf{#1.}}

\definecolor{WowColor}{rgb}{.75,0,.75}
\definecolor{SubtleColor}{rgb}{0,0,.50}

\newcommand{\pathconstraint}{Missed Path Constraint or Sanitization}
\newcommand{\contextmismatch}{Benign Execution Context}
\newcommand{\trustedsource}{Missing Trust Boundary Modeling}
\newcommand{\sinksemantics}{Imprecise Sink Modeling}
\newcommand{\concurrency}{Imprecise Concurrency Modeling}

\newcommand{\qloginj}{Log-injection\xspace}
\newcommand{\qsenslog}{Sensitive-log\xspace}
\newcommand{\qpathinj}{Path-injection\xspace}
\newcommand{\qtaintarith}{Tainted-arithmetic\xspace}
\newcommand{\qxss}{XSS\xspace}
\newcommand{\quserbypass}{User-controlled-bypass\xspace}
\newcommand{\qwidecmp}{Comparison-with-wider-type\xspace}
\newcommand{\qunreleasedlock}{Unreleased-lock\xspace}
\newcommand{\qsqlconcat}{Concatenated-sql-query\xspace}
\newcommand{\qcsrf}{CSRF-unprotected-request-type\xspace}

\newcommand{\RqOneTaxonomyRows}{%
\pathconstraint{} &
CodeQL reports a finding because it misses an explicit or implicit constraint, validation, sanitization, or safe transformation that prevents exploitation. &
Control-flow dependency, 101/203 (49.8\%); explicit validation or sanitization, 64/203 (31.5\%); implicit program invariant, 38/203 (18.7\%) &
183 (36.6\%) \\
\contextmismatch{} &
The alert occurs in code whose execution context limits or removes the relevant attack surface for the analyzed project version. &
Non-production code, 112/177 (63.3\%); attacker-unreachable lifecycle, 26/177 (14.7\%); deployment or bootstrap-limited code, 25/177 (14.1\%); internal runtime component, 14/177 (7.9\%) &
147 (29.4\%) \\
\trustedsource{} &
CodeQL overstates risk because it misses who can control the value, who is authorized to act, or who is able to observe the reported output. &
Authorized or privileged actor, 126/230 (54.8\%); trusted value origin, 76/230 (33.0\%); restricted disclosure channel, 28/230 (12.2\%) &
138 (27.6\%) \\
\concurrency{} &
CodeQL reports a lifecycle or locking risk because it misses cleanup or release behavior in the reviewed context. &
Balanced cleanup, 25/25 (100.0\%) &
25 (5.0\%) \\
\sinksemantics{} &
The reported value reaches a sink, but the sink does not perform the dangerous operation assumed by the query. &
Non-rendering helper or transport sink, 7/7 (100.0\%) &
7 (1.4\%) \\
}

\newcommand{\SelectedRulesRows}{%
\texttt{log-injection} & User input reaches log output & CWE-117 & 11,613 & 31,777 & 99 \\
\texttt{sensitive-log} & Sensitive data reaches logs & CWE-532 & 2,987 & 8,623 & 90 \\
\texttt{path-injection} & User input influences file paths & CWE-022, CWE-023, CWE-036, CWE-073 & 1,195 & 3,567 & 73 \\
\texttt{tainted-arithmetic} & User input affects arithmetic & CWE-190, CWE-191 & 830 & 2,249 & 62 \\
\texttt{xss} & User input reaches web output & CWE-079 & 730 & 1,751 & 60 \\
\texttt{user-controlled-bypass} & User input controls a security check & CWE-807, CWE-290 & 642 & 1,835 & 53 \\
\texttt{comparison-with-wider-type} & Numeric comparison may be misleading & CWE-190, CWE-197 & 460 & 460 & 78 \\
\texttt{unreleased-lock} & Lock may not be released & CWE-764, CWE-833 & 378 & 378 & 50 \\
\texttt{concatenated-sql-query} & User input reaches SQL query text & CWE-089, CWE-564 & 334 & 334 & 37 \\
\texttt{csrf-unprotected-request-type} & Request handler may lack CSRF protection & CWE-352 & 287 & 1,050 & 18 \\
}

\newcommand{\RuleDominantFpReasonRows}{%
comparison-with-wider-type & Implicit program invariant & 35/50 (70.0\%) \\
concatenated-sql-query & Non-production code & 25/50 (50.0\%) \\
csrf-unprotected-request-type & Control-flow dependency & 36/50 (72.0\%) \\
log-injection & Authorized or privileged actor & 22/50 (44.0\%) \\
path-injection & Explicit validation or sanitization & 24/50 (48.0\%) \\
sensitive-log & Restricted disclosure channel & 15/50 (30.0\%) \\
tainted-arithmetic & Explicit validation or sanitization & 25/50 (50.0\%) \\
unreleased-lock & Balanced cleanup & 25/50 (50.0\%) \\
user-controlled-bypass & Control-flow dependency & 48/50 (96.0\%) \\
xss & Authorized or privileged actor & 13/50 (26.0\%) \\
}

\newcommand{\RqTwoRefinementMapRows}{%
Authorized or privileged actor
& Authorized or privileged actor
& G1
& Reviewed privileged, authentication, authorization callables and selected auth decision shapes. \\

Non-production code
& Non-production code
& G0
& Test, example, fixture, benchmark, demo, or generated paths. \\

Control-flow dependency
& Control-flow dependency
& \quserbypass
& Routing or dispatch shaped control flow. \\

Trusted value origin
& Trusted value origin
& G1
& Reviewed callables carrying constants, generated metadata, framework or runtime state, or internal service values. \\

Explicit validation or sanitization
& Explicit validation or sanitization
& G2
& Reviewed validated or sanitized callables plus selected range, clamp, escaping, and normalization patterns. \\

Implicit program invariant
& Implicit program invariant
& \qwidecmp
& Local value, type, or numeric invariants. \\

Restricted disclosure channel
& Restricted disclosure channel
& G4
& Reviewed logging callables for restricted diagnostic, administrative, debug, or internal channels. \\

Attacker-unreachable lifecycle
& Attacker-unreachable lifecycle
& \qunreleasedlock
& Reviewed lifecycle or maintenance lock callables, including startup, cleanup, recovery, and internal background paths. \\

Deployment or bootstrap-limited code
& Deployment or bootstrap-limited code
& \qsqlconcat
& Migration, schema update, installation, deployment, or bootstrap code. \\

Balanced cleanup
& Balanced cleanup
& \qunreleasedlock
& Matching release path, such as unlock in covering \texttt{finally}. \\

Internal runtime component
& Internal runtime component
& G3
& Reviewed runtime, protocol, cache, token/session, metadata, and infrastructure component callables. \\

Non-rendering helper or transport sink
& Non-rendering helper or transport sink
& \qxss
& Internal stream, transport helper, or non-rendering output sink. \\
}

\newcommand{\RqTwoRefinementEffectivenessRows}{%
Authorized or privileged actor & 111/126 (88.1\%) & 526/19154 (2.7\%) \\
Non-production code & 87/112 (77.7\%) & 1620/20000 (8.1\%) \\
Control-flow dependency & 32/50 (64.0\%) & 347/643 (54.0\%) \\
Trusted value origin & 65/76 (85.5\%) & 201/19154 (1.0\%) \\
Explicit validation or sanitization & 61/64 (95.3\%) & 234/14988 (1.6\%) \\
Implicit program invariant & 32/35 (91.4\%) & 114/463 (24.6\%) \\
Restricted disclosure channel & 19/28 (67.9\%) & 85/15056 (0.6\%) \\
Attacker-unreachable lifecycle & 21/26 (80.8\%) & 40/383 (10.4\%) \\
Deployment or bootstrap-limited code & 10/12 (83.3\%) & 18/335 (5.4\%) \\
Balanced cleanup & 20/25 (80.0\%) & 188/383 (49.1\%) \\
Internal runtime component & 10/14 (71.4\%) & 18/16980 (0.1\%) \\
Non-rendering helper or transport sink & 7/7 (100.0\%) & 12/751 (1.6\%) \\
}

\newcommand{\RqThreePerQueryRows}{%
comparison-with-wider-type    & 2 (2) & 3 (3) & 4 (3) & 4 (3) \\
concatenated-sql-query        & 1 (1) & 0 (0) & 5 (4) & 5 (4) \\
csrf-unprotected-request-type & 4 (4) & 2 (2) & 5 (5) & 3 (3) \\
log-injection                 & 1 (1) & 3 (2) & 5 (1) & 5 (1) \\
path-injection                & 0 (0) & 0 (0) & 4 (0) & 5 (3) \\
sensitive-log                 & 0 (0) & 4 (2) & 5 (3) & 4 (2) \\
tainted-arithmetic            & 1 (1) & 0 (0) & 3 (1) & 5 (5) \\
unreleased-lock               & 2 (2) & 3 (3) & 5 (4) & 5 (3) \\
user-controlled-bypass        & 3 (2) & 1 (0) & 5 (3) & 5 (3) \\
xss                           & 1 (1) & 2 (2) & 4 (4) & 5 (4) \\
}

\newcommand{\RqThreePerQueryTotalRow}{%
\textbf{Total} & \textbf{15 (14)} & \textbf{18 (14)} & \textbf{45 (28)} & \textbf{46 (31)} \\
}

\newtcolorbox{takeawaybox}{
  colback=blue!5,
  colframe=blue!25,
  boxrule=0.25pt,
  arc=0pt,
  left=2pt,
  right=2pt,
  top=1pt,
  bottom=1pt,
  before skip=2pt,
  after skip=2pt
}
\begin{document}

\title{An Empirical Analysis of CodeQL False Positives and Query Refinements for Java Vulnerabilities}

\author{
\IEEEauthorblockN{Amirali Sajadi}
\IEEEauthorblockA{
College of Computing \& Informatics\\
Drexel University, Philadelphia, PA, USA\\
amirali.sajadi@drexel.edu
}
\and
\IEEEauthorblockN{Saikat Dutta}
\IEEEauthorblockA{
Department of Computer Science\\
Cornell University, Ithaca, NY, USA\\
saikatd@cornell.edu
}
\and
\IEEEauthorblockN{Preetha Chatterjee}
\IEEEauthorblockA{
College of Computing \& Informatics\\
Drexel University, Philadelphia, PA, USA\\
preetha.chatterjee@drexel.edu
}
}

\maketitle

\begin{abstract}
Static application security testing (SAST) tools help developers find vulnerabilities before deployment. However, false positive reports result in substantial triage efforts. This paper studies whether CodeQL false positives in Java security analysis are recurring, explainable patterns that can be reduced by refining the analysis itself. We run CodeQL’s Java security query suite on 167 CVE instances spanning 110 projects, focusing on the ten queries with highest false positive rates. From these queries, we manually review 500 sampled false positive paths/locations and construct a source-level taxonomy of false positive causes. The five high-level categories of false positive causes are \pathconstraint\ (36.6\%), \contextmismatch\ (29.4\%), and \trustedsource\ (27.6\%), followed by \concurrency\ (5\%) and \sinksemantics{} (1.4\%).

Using insights from our empirical analysis, we implement CodeQL refinements that model recurring false positive patterns to enable their detection and filtering at the query level. Our refinements remove 81.8\% of reviewed false positives. On the full selected-query dataset, they remove 15.8\% of reported paths and locations while retaining 7/8 true positives, showing that many false positives can be reduced directly in the analysis, but also that fixed refinements often depend on project-specific context. To address this generalization gap, we evaluate whether agentic coding tools can adapt refinement patterns to new project-specific contexts.  With our refinement patterns as templates to extend, they succeed on 56\% and 62\% of tasks, respectively, with query compile-pass rates above 90\%. However, without this guidance, the success rates drop to 28\% for both tools, while the compile rate drops to 30-36\%. Overall, our results support a refinement-oriented SAST workflow in which recurring false positives are modeled in CodeQL queries and then automatically adjusted to different project contexts, reducing repeated triage.
\end{abstract}


\section{Introduction}

\begin{figure*}[t]
    \centering
    \includegraphics[width=\textwidth]{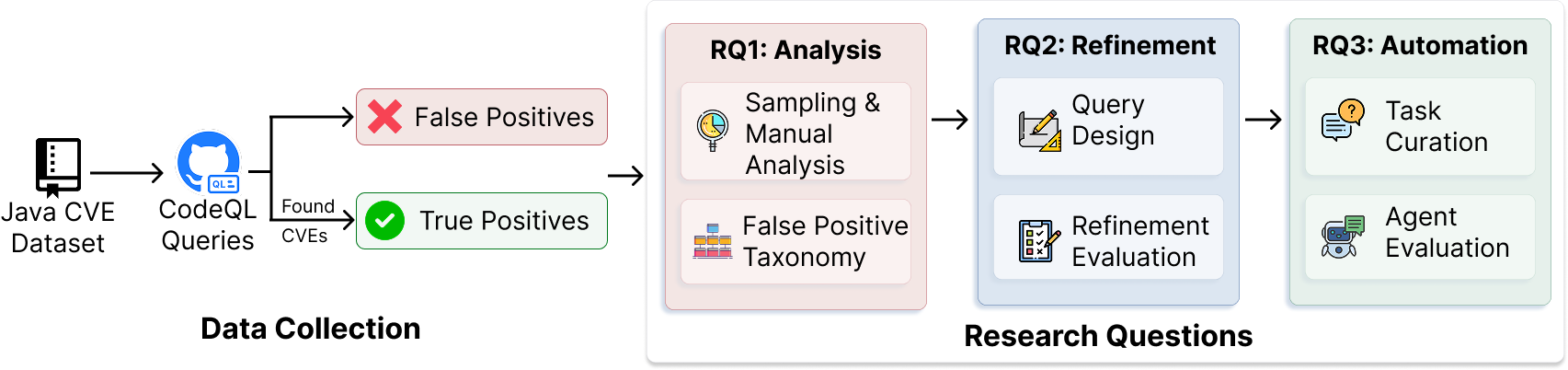}
    \caption{Overview of the Study Methodology.}
    \label{fig:overview}
\vspace{-1.8em}
\end{figure*}

% SAST FP Problem
Static Application Security Testing (or SAST) tools help developers find vulnerabilities before software is deployed. They are valuable as they can be automatically run codebases, can be integrated into CI/CD workflows, and report source locations that developers can inspect for bugs. However, false positive alerts limit their overall effectiveness~\cite{guo2023mitigating, emanuelsson2008comparative, nadeem2012high, medeiros2014automatic, dimastrogiovanni2016towards}. When a tool reports too many alerts that do not correspond to real or exploitable vulnerabilities, developers waste time triaging false alerts instead of fixing real defects. Further, false positives discourage the usage of SAST tools and limit their adoption in broader open-source software. In practice, most developers expect false positive rates below 15--20\% for adoption, while more than 76\% of SAST warnings can be irrelevant to real vulnerabilities~\cite{christakis2016programanalysis, charoenwet2024securereview}.

% Specific FP issues and high-level reason - especially in CodeQL
CodeQL~\cite{codql} is one of the most popular query-based SAST tools. Its query-driven design makes the analysis logic explicit and allows the query to be inspected, modified, and re-executed, making it well suited for systematically studying how and why false positives arise. However, like other SAST tools, CodeQL alerts can still be false positives. CodeQL relies on predefined program patterns and taint analysis specifications to identify potential vulnerabilities. These specifications model how user-controlled data (source) flows through the program and reaches sensitive operations (sinks), but they can often be incomplete, difficult to maintain, outdated, or overly conservative. As a result, benign dataflows may be labeled as dangerous when project-specific logic already ensures security. For example, project-specific authentication mechanisms may not be captured if they are not explicitly modeled in CodeQL's specifications. In these cases, the underlying issue is not only that the alert needs to be filtered, but that the query cannot accurately model the program’s behavior.

Prior work has studied why developers do not adopt SAST and how false positive suppression can be used in practice~\cite{johnson2013staticanalysis,christakis2016programanalysis,hu2025suppressions}. Others have studied the quality of SAST results and taint analysis performance across tools and benchmarks~\cite{cui2024falsepositives,charoenwet2024securereview,li2024evaluating,luo2019qualitative,arzt2015using,qiu2018analyzing}.  However, these studies do not use source-level false positive evidence to refine the underlying SAST analyses, nor do they study whether such empirically grounded refinements can be adapted to project-specific contexts. Recent work has also used post-processing, ranking, exploit generation, and LLM-based workflows to reduce false positives~\cite{li2025iris,wei2017oasis,wen2024automatically,wang2025qlcoder}. However, LLM-based methods need to be used after every run, which can become very expensive very quickly in CI and they are also prone to imprecise reasoning, which may be hard to reproduce. This motivates the question of how to systematically incorporate fine-grained, project-specific evidence into SAST analyses without incurring high manual and monetary cost.

% Our work
In this paper, we present a systematic study of false positive reduction through CodeQL query refinements. Prior false positive studies mostly characterize static-analysis warnings, study developer reactions, or filter alerts after they are reported. Our study instead connects source-level false positive causes to actionable changes in the CodeQL queries themselves. Grounded in our empirical analysis, we design refinements that model recurring false positive patterns directly in CodeQL query logic. We then study whether agentic LLMs can adapt these refinements to project-specific contexts. This differs from LLM-based false positive filtering, where the model typically filters the alerts after analysis has already run. In our setting, the final artifact is a modified CodeQL query that can be inspected, rerun, and reused to prevent similar false positives in future analyses. However, maintaining separate refinements for each project can be costly, since codebases and threat models change over time. Our approach therefore studies reusable refinements and agent-assisted adaptation as a way to make project-specific query refinement more practical.

% RQ1 finds the source-level reasons needed for refinement.
We frame our study around three main research questions:

\textit{\noindent\textbf{RQ1: What are the source-level reasons behind CodeQL false positives in Java projects?}}
First, we run CodeQL's `security-extended' query suite on 167 CVEs, focusing on the top ten queries that produce the most false positives. We then manually inspect a statistically significant subset of 500 confirmed false positives and identify fine-grained causes grouped into five broader families, i.e., \pathconstraint, \contextmismatch, \trustedsource, \concurrency, and \sinksemantics. More importantly, this analysis identifies the patterns behind the project-specific refinements which must be modeled to adequately address false positives in CodeQL queries.

% RQ2 shows that many false positives can be removed by CodeQL refinements.
\textit{\textbf{RQ2. Can targeted query refinements reduce recurring false positive patterns?}}
We use the insights from RQ1 to design CodeQL refinements that encode recurring false positive causes. Our refinements remove 409/500 (81.8\%) reviewed false positives, retaining 7/8 (87.5\%) true positives across the full dataset. While developers can suppress individual CodeQL alerts at specific source locations, these suppressions only affect the singular instance and do not address the underlying analysis pattern. In contrast, our refinements address the recurring false positive patterns directly in the query logic.

% RQ3 finds that agents can generalize our refinements (but not write them from scratch)
\textit{\noindent\textbf{RQ3: Can agentic LLMs adapt CodeQL refinements to project-specific contexts and prevent future false positives?}}
Though our RQ2 refinements significantly reduce false positives within the reviewed sample, their generalizability remains limited as many refinements are project-specific. This motivates our study of agent-based query adjustments. A developer may review an alert and decide it is a false positive in a specific project. We ask whether this feedback can be used to adapt the relevant RQ2 refinement to that project. In this setting, the agent receives a false positive, the project context, and the corresponding refinement queries from RQ2 as a template to extend. The agent then adjusts the query in order to eliminate the project-specific false positive pattern. We evaluate this workflow using OpenAI Codex (GPT 5.5) and Claude Code (Claude Opus 4.6) agents, showing that guided refinements provide evidence-based reductions in false positives in up to 56\% of cases, compared to 28\% when agents write refinements without our RQ2 templates. The large gap between guided and unguided runs highlights the value of empirical refinement patterns as reusable templates for project-specific query adaptation.

Our study reveals the following novel insights: First, our results demonstrate that an overwhelming majority of false positives can be captured using source-level patterns. Second, most of these patterns can be modeled via CodeQL via targeted query refinements, which drastically reduce false positives while retaining true positive alerts. Third, these refinements are very project-specific and so, by construction, they are very hard to easily generalize to many projects. Finally, our early evaluation of LLM agents reveal that they could be useful in automating the query refinement task to some extent but they often struggle with understanding the semantics of CodeQL and the code pattern. So, more future research is needed to carefully guide LLM agents into automating these refinements. Our full replication package is available at \cite{codeql_fp_study_replication}.

\section{Methodology}
\label{sec:methodology}

Figure~\ref{fig:overview} shows the overview of our study design. We use a recent benchmark, CWE-Bench-Java~\cite{li2025iris}, that contains 213 CVEs across 135 Java projects (Section~\ref{sec:methodology-dataset}). To answer RQ1, we run CodeQL and analyze the resulting alerts to characterize the root causes of false positives (Section~\ref{sec:methodology-rq1}). For RQ2, we use the RQ1 taxonomy to design targeted CodeQL query refinements and evaluate how many reviewed false positives they remove while preserving the true positives (Section~\ref{sec:methodology-rq2}). For RQ3, we examine whether coding agents can adapt these refinements to new project-specific contexts (Section~\ref{sec:methodology-rq3}). 

\begin{table*}[t]
\centering
\footnotesize
\setlength{\tabcolsep}{3pt}
\renewcommand{\arraystretch}{1.08}
\caption{Selected CodeQL queries ranked by eligible false positive count.}
\label{tab:selected-rules}
\resizebox{\textwidth}{!}{%
\begin{tabular}{lllrrr}
\toprule
Query & Description & CodeQL CWE tags & FP findings & Paths and locations & Projects \\
\midrule
\SelectedRulesRows
\bottomrule
\end{tabular}%
}
\vspace{-2em}
\end{table*}

\subsection{Dataset and Sampling}
\label{sec:methodology-dataset}

\noindent\textbf{Dataset and CodeQL results.}
Our study uses Java CVE projects from the CWE-Bench-Java dataset~\cite{li2025iris}. The benchmark contains 213 high-severity Java vulnerabilities across 45 CWE classes, including path traversal (CWE-022), cross-site scripting (CWE-079), command injection (CWE-078), code injection (CWE-094), XXE (CWE-611), and authentication errors (CWE-287). Each CVE instance includes vulnerability metadata, the vulnerable and fixed project versions, the code locations changed by the security fix, and scripts to compile each project version with CodeQL. The dataset also provides containerized vulnerable versions that can be built and analyzed. This lets us run CodeQL and inspect alerts, paths, and locations in the same project context. 

We run CodeQL v2.25.4 on each vulnerable project version using the Java \texttt{security-extended} suite from \texttt{codeql/java-queries} v1.11.2. The resolved suite includes 124 queries. CodeQL produces one SARIF file per run, containing the alerts reported by these queries.
167 vulnerable project versions successfully built and produced CodeQL reports across 110 unique projects. We use the same procedure as \cite{li2025iris} to classify each alert as a true positive or false positive: an alert that contains path(s) triggering the vulnerable methods related to the CVE is considered a true positive. While the remaining paths can also indicate true positives, we defer this reasoning to manual analysis in RQ2~\ref{sec:results-rq2}.

\noindent\textbf{Sampling strategy.}
We sample false positives from the queries that create the largest triage burden. We use alert count rather than path count to rank the frequency of queries because developers first encounter a CodeQL report as alerts, even when one alert contains multiple reported paths. We then select the ten queries with the most false positive alerts.

For manual review, we sample the concrete code evidence reported by CodeQL. For path-producing alerts, each SARIF \texttt{codeFlows} entry provides one source-to-sink path that can be sampled. For location-only alerts, the primary reported location provides one location that can be sampled. Thus, the sampling pool for each query contains the paths and locations reported by that query.

We use random sampling across the ten selected queries. For each query, we randomly sample 50 eligible paths or locations from that query's pool, yielding 500 false positives for manual review. This balanced selection gives each query enough reviewed examples for query-level characterization.

The selected queries contain 19,456 eligible false positive alerts, which expand to 52,024 eligible paths and locations combined. Using the standard finite population sample size formula with 95$\pm$5\% confidence, this population requires 382 samples~\cite{rv1970determining, bujang2017simplified, sajadi2023towards, shu2026empirical}. Our 500 instances exceed this requirement and provide a statistically representative sample of the selected false positive paths and locations.

\subsection{RQ1: False Positive Characterization}
\label{sec:methodology-rq1}

\noindent\textbf{Review artifacts.}
For RQ1, we manually review each sampled false positive review unit using a structured source context artifact. The artifact is generated from the CodeQL SARIF output and the corresponding vulnerable project source tree. For each reported location, our scripts map the file and line number to the enclosing Java method or constructor declaration. For path producing results, the artifact also contains the ordered CodeQL source to sink path, the file and line of each node, and the enclosing declaration for the source, sink, and intermediate nodes when available.

During our review process, these artifacts provided the starting context for each instance. When the reported path did not provide enough context, we inspected the full vulnerable project. Notably, the artifact renders relevant code snippets with inline hop annotations. Reviewers use these hops to follow the reported source-to-sink flow and inspect the surrounding code when needed.

\begin{comment}
\begin{lstlisting}[style=javacode, language=Java,basicstyle=\ttfamily\scriptsize]
public void channelRead(ChannelHandlerContext ctx, Object msg) { // hop 0
processReceived(ctx.channel(), (Command) msg); // hop 1
}

private void processReceived(final Channel channel,
final Command command) { // hop 2
processByCommandType(channel, command); // hop 3
}

public void processByCommandType(final Channel channel,
final Command command) { // hop 4
logger.warn("receive response {}, but not matched any request",
command); // hop 5
}
\end{lstlisting}
\end{comment}

\noindent\textbf{Manual review process.}
We use inductive coding~\cite{saldana2021coding, braun2006using}, i.e., we derive labels from observations during our review to identify recurring patterns of false positives in the dataset. We inspect each sampled path or location and assess whether it could correspond to the vulnerability reported by the CodeQL query in that project.

Throughout the review, recurring distinctions became visible around execution context, source control, trust boundaries, value flow, guards, and sink behavior. We record five pieces of evidence for each instance: 1) we identify the \textit{execution context}, such as production, test, tooling, framework, administrative, or internal cluster code; 2) we identify\textit{ who could control the reported source value} in practice e.g., an unauthenticated user, privileged user, internal node, test harness, etc.; 3) we record what \textit{kind of value was controlled}, such as a path, SQL fragment, identifier, configuration value, or message object; 4) we check \textit{whether the reported value reached the sink} as data or only affected control flow, and whether the sink was security relevant in its runtime context; and 5) we record any \textit{protection before the sink}, including authorization, sanitization, allowlisting, parameterization, or existence checks. After review, we write an explanation of the root cause. For example, a path-injection alert may be marked as having adequate sanitization when the reported path data is first resolved to an absolute path and then checked against an allowed base directory before the file operation.

\noindent\textbf{Root cause labeling and taxonomy construction.}
We construct a hierarchical taxonomy of false positives through iterative coding. For each reviewed false positive, we assign one or more fine-grained root cause labels and mark one label as primary. The primary label captures the false positive root cause that most directly explains why the reported CodeQL path or location does not correspond to an exploitable vulnerability. When multiple facts contribute to the judgment, we record them as secondary labels or notes.

We then review the written explanations for each false positive, group recurring explanations into fine-grained labels, and refine label definitions. After finalizing the fine-grained labels, we group related labels into five higher-level families. For example, labels involving explicit validation, constrained control flow, and implicit program invariants are grouped under the path constraint family.

\subsection{RQ2: Query Refinement for False Positive Reduction}
\label{sec:methodology-rq2}

\noindent\textbf{Refinement design.}
RQ2 evaluates whether the fine-grained false positive mechanisms from RQ1 can be translated into CodeQL refinements. For each fine-grained root-cause category, we design a refinement predicate that captures the program evidence used during manual review. These predicates encode recurring evidence such as file paths, enclosing callables, control-flow structure, validation and sanitization patterns, value invariants, trusted value origins, authorization contexts, logging contexts, lifecycle code, and lock cleanup structure. The goal of RQ2 is not to produce a universal replacement for the original CodeQL queries, but to test whether query refinements can model the underlying issues detected in RQ1 and how much false positive reduction those refinements provide while preserving baseline true positives i.e., paths/locations that match the benchmark CVE using the CWE-Bench-Java labeling procedure described above.

Each refinement is implemented as a modified version of an existing CodeQL query in Table~\ref{tab:selected-rules}. The modified query reuses the original logic and adds suppression predicates for specific false positive patterns. A result is reported only if it is produced by the original query and does not match the refinement predicates. For each candidate refinement, we define four elements before implementation. \textit{First}, we identify the false positive pattern from the RQ1 taxonomy. \textit{Second}, we decide which CodeQL rules can be refined to address that pattern. \textit{Third}, we identify the evidence needed to recognize the pattern, such as source file paths, enclosing callables, dataflows, expression names, types, annotations, packages, or method calls. \textit{Fourth}, we choose the suppression point. Depending on the rule, the predicate may apply to the source, sink, enclosing callable, primary expression, or reported method.

We write the predicates conservatively. A refinement suppresses a finding only when the code contains evidence for the false positive pattern. If the same pattern could hide a real vulnerability, we make the predicate more restrictive. For example, a source-trust refinement may suppress logging of application-generated identifiers such as request IDs, operation names, or status values. It does not suppress values whose names suggest sensitive data, such as passwords, tokens, credentials, or secrets. Similarly, a sink-semantics refinement may suppress an XSS finding only when the value is passed through a known escaping or framework rendering API. It does not suppress arbitrary writes to an HTTP response.

\noindent\textbf{Evaluation.}
We evaluate each refinement at two levels. First, we apply it to the 500 manually reviewed false positives. This checks whether the refinement removes the RQ1 patterns it was designed to target. Second, we apply it to all applicable false positives outside the reviewed sample, checking whether the pattern generalizes beyond the manually reviewed 500. We consider a false positive removed when it appears in the original CodeQL output but is suppressed by the refined query. We also check whether any true positives are removed, measuring the impact of our refinements on precision.

\subsection{RQ3: Query Refinement Using LLM Agents}
\label{sec:methodology-rq3}

\noindent\textbf{Task construction.}
To answer RQ3, we construct 50 project-specific refinement task instances from false positives outside the 500 instances used in RQ1. A task instance is a concrete query-refinement problem. It contains a target false positive, the vulnerable project source code, and the corresponding CodeQL query that reported the false positive path or location. We use the target false positive as developer feedback, representing a case where a developer has reviewed a CodeQL-reported path or location and marked it as a false positive for that project. The agent's task is to modify the CodeQL query so that it suppresses the project-specific false positive pattern without removing known true positives. Our task sample comes from the broader dataset after excluding the 500 false positives used in RQ1. For each modified CodeQL query, we sample 5 additional false positives, yielding 50 task instances.

For each task instance, we create two agent runs. Both runs receive the task instance as input. The guided run additionally receives the relevant RQ2 refinement pattern, defined as the general query logic developed in RQ2 for the same query and false positive reason. The agent must extend or adapt this pattern to the local context of the project. The unguided run does not receive the RQ2 refinement pattern and must design the CodeQL refinement on its own. This comparison evaluates whether the evidence-based RQ2 refinements provide useful guidance for project-specific adaptation.

The guided agent receives only the relevant RQ2 refinement pattern as a template, not a task-specific refined query for the target false positive. The unguided agent receives no RQ2 refinement pattern. In both settings, the agent must produce a semantic query-level refinement rather than an alert-specific suppression, and we reject submissions that remove the target only by matching the sampled file path, line number, result index, or project name. All prompts are available in our replication package~\cite{codeql_fp_study_replication}.

\begin{table*}[t]
\centering
\scriptsize
\setlength{\tabcolsep}{3pt}
\renewcommand{\arraystretch}{1.14}
\caption{False-positive taxonomy with primary group counts and within-group fine-grained breakdowns. Fine-grained counts include primary and secondary labels, so their within-group totals may exceed the primary count for the group.}
\label{tab:fp-taxonomy-updated}
\begin{tabular}{p{0.19\textwidth}p{0.35\textwidth}p{0.32\textwidth}r}
\toprule
\textbf{FP group} &
\textbf{Definition} &
\textbf{Fine-grained breakdown} &
\textbf{Primary FPs} \\
\midrule
\RqOneTaxonomyRows
\bottomrule
\end{tabular}
\vspace{-2.5em}
\end{table*}

\noindent\textbf{Evaluation.}
We use the following process to evaluate the guided and unguided runs. First, we rerun the submitted query on the target project database and check whether it removes the false positive used as feedback. Second, we run the submitted query against the full selected-query dataset and check whether it removes any known true positives. This second check measures whether the project-specific adaptation sacrifices recall while suppressing the target false positive.

We selected Codex with GPT 5.5~\cite{openai_codex_cli, openai_gpt55} and Claude Code with Claude Opus 4.5~\cite{anthropic_claude_code, anthropic_claude_opus_45} as our agentic coding systems. The task requires editing CodeQL queries in a repository, interpreting project context, and running commands, which makes agentic tool use more relevant than prompt-based code generation. Additionally, both these agents have proven to be amongst the highest performing agentic systems~\cite{agent_arena_leaderboard}.

\section{Results}
\newcommand{\category}{category\xspace}
\newcommand{\categories}{categories\xspace}

\subsection{RQ1: False Positive Reason Taxonomy}
\label{sec:results-rq1}

Our RQ1 dataset contains 500 manually confirmed false positives, each labeled with a primary fine-grained \category that explains why the reported CodeQL vulnerability does not correspond to a real vulnerability in context, along with secondary \categories when additional factors contribute to the judgment.
For example, a path may primarily be a false positive because the tainted value is explicitly sanitized, while also involving an operation restricted to an authorized or privileged actor. In such cases, the primary \category is the reason we judge to be the most direct explanation for the false positive, e.g., sufficient sanitization.

Overall, the fine-grained \categories fall into five higher-level false positive groups, as summarized in
Table~\ref{tab:fp-taxonomy-updated}. The group-level counts use the primary \category assigned to each false positive, so the primary counts sum to 500. The fine-grained breakdowns include both primary and secondary \categories, so those totals can be larger than the group-level primary count.

We observe that the most prevalent reason of FPs is \pathconstraint{} (183/500, 36.6\%), followed by \contextmismatch{} (147/500, 29.4\%) and \trustedsource{} (138/500, 27.6\%). Other categories include \concurrency{} (25/500, 5.0\%) and \sinksemantics{} (7/500, 1.4\%). Across these FP groups, the CodeQL query usually matched the pattern it was designed to find, but manual review identified additional factors explaining why these patterns do not constitute real vulnerabilities. These factors include the effective trust or disclosure boundary of the reported value, the environment in which code is executed, checks or transformations applied before a sink, and the behavior of the relevant API. Next, we describe the five high-level FP groups in detail. 

\newcommand{\query}{query\xspace}
\newcommand{\queries}{queries\xspace}

\begin{enumerate}[
  leftmargin=0pt,
  labelwidth=0pt,
  labelsep=0.4em,
  itemindent=1.5em,
  topsep=0pt,
  itemsep=0pt,
  parsep=0pt
]
\item\textbf{\pathconstraint{}.}
The most frequent group covers alerts where a relevant flow, value, or condition exists, but the reviewed code prevents the reported behavior from being exploitable. Fine-grained reasons in this group include explicit validation or sanitization, i.e., checks or transformations before the sink; implicit program invariants, such as numeric bounds or identifier formats that make the reported behavior infeasible; and control-flow dependency which refers to request-influenced routing rather than attacker-controlled data reaching a dangerous sink.

Listing~\ref{lst:path-constraint-example} shows a path-injection false positive. CodeQL reports that a dataset name derived from the HTTP request path reaches \texttt{IO.deleteAll()}, a recursive filesystem deletion operation. The path is a false positive because the value is constrained before deletion: it is converted into a dataset name in lines 1--2, checked against the data access point registry in lines 4--8, resolved under Fuseki's configured database root in lines 14--17, and deleted only after existence and symbolic-link checks in lines 19--21. Together, these checks show that the value is not used as an attacker-chosen file path. Similar constraints appear in user-controlled-bypass and csrf-unprotected-request-type, where request-influenced branches or endpoint choices are constrained by the code.

\begin{lstlisting}[
  style=javacode,
  caption={Example of \pathconstraint{} in apache/jena.},
  label={lst:path-constraint-example}
]
String name = DataAccessPoint.canonical(
    getItemName(action));

if (!action.getDataAccessPointRegistry()
        .isRegistered(name)) {
    ServletOps.errorNotFound(
        "No such dataset registered: " + name);
}

DataAccessPoint ref =
    action.getDataAccessPointRegistry().get(name);
action.getDataAccessPointRegistry().remove(name);

String filename = name.startsWith("/")
    ? name.substring(1) : name;
Path pDatabase =
    FusekiWebapp.dirDatabases.resolve(filename);

if (Files.exists(pDatabase)
        && !Files.isSymbolicLink(pDatabase)) {
    IO.deleteAll(pDatabase);
}
\end{lstlisting}
\vspace{-1em}

\item\textbf{\contextmismatch{}.}
The second largest group contains alerts in code that is outside the relevant production attack surface of the analyzed project version. At the fine-grained level, this group includes non-production code such as tests, examples, fixtures, and evaluation utilities; deployment or bootstrap-limited code such as setup or migration logic; and attacker-unreachable lifecycle code such as framework paths outside the relevant production attack surface.
Listing~\ref{lst:execution-context-example} shows a SQL-concatenation false positive from apache/tika. CodeQL reports a SQL constructed by concatenating strings, executed through \texttt{Statement.executeUpdate()}. The pattern is relevant, but the surrounding code belongs to Tika's offline evaluation tooling. It reads parser error logs from files supplied to the evaluation command and updates an evaluation database. The reviewed context does not expose this SQL construction to untrusted production input.

\begin{lstlisting}[
  style=javacode,
  caption={Example of \contextmismatch{} in apache/tika.},
  label={lst:execution-context-example}
]
// Offline evaluation tool.
public static void main(String[] args) throws Exception {
    Path xmlLogFileA = Paths.get(args[0]);
    Path db = Paths.get(args[2]);

    Connection connection = new H2Util(db).getConnection();
    writer.update(connection,
        ExtractComparer.EXTRACT_EXCEPTION_TABLE_A,
        xmlLogFileA);
}

private void update(String errorTableName, String filePath)
        throws SQLException {
    String sql = "UPDATE " + errorTableName
        + " SET " + Cols.FILE_PATH + "='" + filePath + "'"
        + " where " + Cols.CONTAINER_ID + "="
        + getContainerId(filePath);

    statement.executeUpdate(sql);
}
\end{lstlisting}

\item\textbf{\trustedsource{}.}
\trustedsource{} false positives arise when CodeQL treats a value as attacker-controlled or attacker-observable, but project context shows that the value is constrained by authorization, privileged control, trusted origin, or limited disclosure, i.e., restricted log/output access. At the fine-grained level, authorized or privileged actor covers administrator-only or internal-service workflows, trusted value origin covers values produced by the application, framework, cluster peers, or deployment configuration, and restricted disclosure channel covers outputs limited to logs or internal channels.
Listing~\ref{lst:source-trust-example} shows a path-injection false positive from apache/kylin. CodeQL reports an HTTP request-supplied value flowing into \texttt{FileUtils.copyFile()} in line 17, where the destination file uses the request-derived principal value. The reported dataflow is accurate: a value from the request influences the file path. We classify the alert as a false positive because the operation is part of an administrator-only workflow. The \texttt{@PreAuthorize} annotation in lines 1--2 requires either a global administrator role or project-level administration permission before the method can execute. This authorization boundary places the operation outside the untrusted user attack surface. The path-injection query does not make that project-specific threat-model judgment; it treats request-controlled filesystem access as a vulnerability even when the request reaches the sink only through a privileged administrative operation.

\begin{lstlisting}[
  style=javacode,
  caption={Example of \trustedsource{} in apache/kylin.},
  label={lst:source-trust-example}
]
@PreAuthorize(Constant.ACCESS_HAS_ROLE_ADMIN
    + " or hasPermission(#project, 'ADMINISTRATION')")
public void updateProjectKerberosInfo(
        String project,
        ProjectKerberosInfoRequest request) throws Exception {
    checkAndReplaceProjectKerberosInfo(project,
        request.getPrincipal());
    backupAndDeleteKeytab(request.getPrincipal());
}

public File backupAndDeleteKeytab(String principal)
        throws IOException {
    File kFile = new File(
        KapConfig.getKylinConfDirAtBestEffort(),
        principal + KEYTAB_SUFFIX);

    FileUtils.copyFile(kTempFile, kFile);
    return kFile;
}
\end{lstlisting}

\item\textbf{\concurrency{}.}
The concurrency group covers alerts where CodeQL reports a lock or lifecycle operation as potentially unbalanced. Unlike the preceding groups, these reports are not about attacker-controlled values reaching security-sensitive sinks. Instead, they depend on whether the analysis accurately tracks acquisition and release behavior across control-flow paths.
Listing~\ref{lst:concurrency-lifecycle-example} shows an unreleased-lock false positive from apache/rocketmq. CodeQL reports that the write lock acquired through \texttt{lockInterruptibly()} might not be unlocked, or might be locked more times than it is unlocked. In the reviewed method, however, the acquisition appears inside an inner \texttt{try} block whose corresponding \texttt{finally} block releases the write lock before control can leave the protected region. The method is also wrapped by an outer \texttt{catch}; under Java's \texttt{try}/\texttt{finally} semantics, exceptions from the protected region run the \texttt{finally} block before reaching that outer handler. The alert is therefore best explained as a conservative control-flow/lifecycle abstraction; the query finds a path whose total lock count remains positive at method exit, even though manual review shows that the relevant path is covered by the \texttt{finally} cleanup.

\begin{lstlisting}[
  style=javacode,
  caption={Example of \concurrency{} in RocketMQ.},
  label={lst:concurrency-lifecycle-example}
]
public void deleteTopic(
        String topic, String clusterName) {
    try {
        try {
            this.lock.writeLock()
                .lockInterruptibly();
            // update topic routing tables
        } finally {
            this.lock.writeLock().unlock();
        }
    } catch (Exception e) {
        log.error("deleteTopic Exception", e);
    }
}
\end{lstlisting}

\item\textbf{\sinksemantics{}.}
Some alerts reach APIs that match a query's sink, but the concrete API behavior does not create the security risk modeled by the query. At the fine-grained level, non-rendering helper or transport sink covers internal byte movement rather than browser-rendered output, while framework or template escaped sink covers output that passes through rendering or escaping behavior before reaching the client.
Listing~\ref{lst:sink-semantics-example} shows an XSS false positive from Apache apache/nifi. CodeQL reports data from a socket input stream flowing into a write operation. However, the concrete write target is an internal stream wrapper used while transferring NiFi content, not an HTTP response writer or template renderer. The write copies bytes through NiFi's internal content/protocol path and does not create the browser HTML or JavaScript interpretation for XSS.

\begin{lstlisting}[
  style=javacode,
  caption={Example of \sinksemantics{} in Apache NiFi.},
  label={lst:sink-semantics-example}
]
// SocketInput.java
public SocketInput(final Socket socket)
        throws IOException {
    this.socket = socket;
    socketIn = socket.getInputStream();
    countingIn = new ByteCountingInputStream(socketIn);
    bufferedIn = new BufferedInputStream(countingIn);
    interruptableIn =
        new InterruptableInputStream(bufferedIn);
}

// TaskTerminationOutputStream.java
public void write(byte[] b, int off, int len)
        throws IOException {
    verifyNotTerminated();
    delegate.write(b, off, len);
}
\end{lstlisting}
\end{enumerate}
\vspace{-.4em}

\begin{table}[t]
\centering
\scriptsize
\setlength{\tabcolsep}{2.2pt}
\renewcommand{\arraystretch}{1.06}
\caption{Dominant primary fine-grained false positive category by selected CodeQL \query{}. Each \query{} contributes 50 confirmed false positives. \emakefirstuc{\query{}} names omit the java/ prefix.}
\label{tab:rule-dominant-fp-reasons}

\begin{tabular}{p{0.42\columnwidth}p{0.40\columnwidth}c}
\toprule
\textbf{\emakefirstuc{CodeQL \query{}}} & \textbf{Dominant \category} & \textbf{FP rate} \\
\midrule
\RuleDominantFpReasonRows
\bottomrule
\end{tabular}
\vspace{-2.7em}
\end{table}

\mypara{Distribution of Dominant Categories across CodeQL Queries}
Table~\ref{tab:rule-dominant-fp-reasons} shows that dominant false positive reasons vary substantially across CodeQL \queries{}. For user-controlled-bypass and csrf-unprotected-request-type, the dominant issue is control-flow dependency: the alert is usually tied to a request-influenced branch or endpoint choice, but manual review does not find attacker-controlled data bypassing a concrete security decision. Other \queries{} depend more on value constraints and validation, as in tainted-arithmetic and path-injection, or implicit invariants, as in comparison-with-wider-type. Logging-related \queries{} depend more on actor capability, trusted value origin, or restricted disclosure of the reported output. XSS findings are more diffuse after the sink-semantics audit, with authorized or privileged actor context forming the largest single category. This \query{}-level variation motivates RQ2, where refinements must be tailored to specific false positive \categories{} rather than applied as a generic filter.

\begin{takeawaybox}
\textbf{RQ1 Summary:} CodeQL false positives follow recurring source-level patterns, but dominant patterns vary by query, suggesting that refinements should be query-specific.
\end{takeawaybox}

\subsection{RQ2: Effect of Query-Level False Positive Refinements}
\label{sec:results-rq2}

\begin{table}[t]
\centering
\scriptsize
\setlength{\tabcolsep}{3pt}
\renewcommand{\arraystretch}{1.06}
\caption{Effectiveness of RQ2 refinements on reviewed false positives and the full selected-query dataset. The total row de-duplicated removals across refinements.}
\label{tab:rq2-refinement-effectiveness}
\begin{tabularx}{\columnwidth}{p{0.45\columnwidth}p{0.24\columnwidth}p{0.24\columnwidth}}
\toprule
\textbf{Target RQ1 reason} &
\textbf{Target FP rem.} &
\textbf{Full FP rem.} \\
\midrule
\RqTwoRefinementEffectivenessRows
\midrule
\textbf{De-duplicated total} & \textbf{409/500 (81.8\%)} & \textbf{3158/20000 (15.8\%)} \\
\bottomrule
\end{tabularx}
\vspace{-1.8em}
\end{table}

\begin{table*}[t]
\centering
\scriptsize
\setlength{\tabcolsep}{3pt}
\renewcommand{\arraystretch}{1.08}
\caption{Reported RQ2 refinements and their target RQ1 false positive reasons. Rule groups are defined below the table.}
\label{tab:rq2-refinement-map}
\begin{tabularx}{\textwidth}{p{0.22\textwidth}p{0.22\textwidth}p{0.12\textwidth}X}
\toprule
\textbf{Target RQ1 reason} &
\textbf{RQ2 refinement} &
\textbf{Applied rule(s)} &
\textbf{Predicate condition} \\
\midrule
\RqTwoRefinementMapRows
\bottomrule
\vspace{1pt}
\end{tabularx}

\raggedright
\scriptsize
G0: all ten studied \queries{}.
G1: \qloginj, \qsenslog, \qpathinj, \qtaintarith, \qxss, \quserbypass, \qsqlconcat, \qcsrf.
G2: \qsqlconcat, \qloginj, \qpathinj, \qtaintarith, \qxss.
G3: \qsqlconcat, \qloginj, \qsenslog, \qtaintarith, \qxss.
G4: \qloginj, \qsenslog.
\vspace{-2.7em}
\end{table*}

RQ2 evaluates whether fine-grained false positive mechanisms identified in RQ1 can be used to create CodeQL query refinements. We implement refinements for the fine-grained mechanisms in the 500 manually reviewed false positives whose evidence can be expressed as query-level predicates. This evidence includes enclosing callables, local control-flow structure, query-construction patterns, value invariants, authorization checks, and framework-specific contexts. The refinements test whether the program facts used during manual false positive review can also be modeled inside CodeQL. RQ2 therefore measures how many reviewed false positives these refinements remove, whether they preserve true positives, and how many true/false positives they remove in the full dataset.

\noindent\textbf{Baseline true positive counts.}
Before evaluating refinements, we first measure how often the baseline CodeQL run identifies known benchmark vulnerabilities covered by the selected rules. CodeQL detected 7 of 81 applicable benchmark vulnerabilities (8.6\%), yielding 56 true positives. These true positives came from two rules. \qpathinj{} detected 5 of 52 applicable vulnerabilities and produced 53 true positives, while \qxss{} detected 2 of 24 applicable vulnerabilities and produced 3 true positive paths. \qsenslog{} had 3 applicable benchmark vulnerabilities and detected none, while \qsqlconcat{} had 2 applicable benchmark vulnerabilities and detected none. The remaining selected rules had no applicable benchmark vulnerability in this dataset. We use these 56 true positives to observe whether the refinements preserve or remove them.

\noindent\textbf{Refinement scope and effectiveness.}
Table~\ref{tab:rq2-refinement-map} maps each refinement to the RQ1 false positive reason it addresses. The refinements cover all fine-grained mechanisms in the RQ1 taxonomy. The table summarizes the rule scope and the predicate condition used by each refinement.

% \noindent\textbf{Refinement effectiveness.}
Table~\ref{tab:rq2-refinement-effectiveness} reports both targeted and full dataset effectiveness. Target false positive removal measures whether each refinement removes false positives within the 500 reviewed instances, while full false positive removal measures reduction across all in-scope results in the dataset.

Overall, the refinements remove 409/500 (81.8\%) reviewed false positives after de-duplication. They also remove 3158/20000 (15.8\%) reported paths/locations in the full selected-query dataset. The strongest targeted reductions occur for non-rendering helper or transport sink (7/7, 100.0\%), explicit validation or sanitization (61/64, 95.3\%), implicit program invariant (32/35, 91.4\%), authorized or privileged actor (111/126, 88.1\%), and trusted value origin (65/76, 85.5\%). The remaining reviewed false positives often depend on runtime or configurations, such as deployment modes, framework configurations, or threat models assumptions. At the same time, the refinements preserve most known true positives. After de-duplicating overlaps across refinements, they retain 49/56 (87.5\%) baseline true positives. The seven removed true positives come from non-production code and from overlapping trusted value origin and explicit validation or sanitization predicates.

\noindent\textbf{Refinement Mechanisms.} All refinements fall into one of the following implementation styles. The listings show excerpts from the refinement query files, focusing on predicates used to suppress recurring false positive patterns. The full refinement queries are included in our replication package~\cite{codeql_fp_study_replication}.

\textit{\underline{The first style}} suppresses false positives using syntactic or structural evidence that CodeQL can recognize locally. This includes non-production paths, deployment or bootstrap code, balanced lock cleanup, and non-rendering helper or transport sinks. The non-rendering sink predicate removes the NiFi \sinksemantics{} example in RQ1, where the reported write is an internal content/protocol stream rather than browser-rendered output. These refinements are comparatively direct because the predicate can match file paths, enclosing methods, sink wrappers, or release paths without requiring a project-wide threat model. Listing~\ref{lst:rq2-cleanup-refinement} shows a simplified balanced-cleanup example that suppresses an \qunreleasedlock{} reported location when the same lock receiver is released in a covering \texttt{finally} block. This refinement removes the RocketMQ \concurrency{} example in RQ1.

\begin{lstlisting}[
  language=CodeQL,
  style=codeqlstyle,
  caption={Simplified structural refinement for unreleased-lock false positives.},
  label={lst:rq2-cleanup-refinement}
]
import java
import semmle.code.java.Concurrency

/** Refinement: same receiver, not only same lock type. */
predicate sameLockReceiver(Expr a, Expr b) {
  a.(VarAccess).getVariable() = b.(VarAccess).getVariable()
  or
  exists(FieldAccess fa, FieldAccess fb |
    fa = a and fb = b and fa.getField() = fb.getField()
  )
}

/** Refinement: acquisition has same-lock cleanup. */
predicate hasBalancedCleanup(MethodCall lock) {
  exists(LockType t, MethodCall unlock |
    lock = t.getLockAccess() and
    unlock = t.getUnlockAccess() and
    sameLockReceiver(lock.getQualifier(), unlock.getQualifier()) and
    unlock.getEnclosingStmt().getParent*() instanceof FinallyStmt
  )
}

from MethodCall lock
where not hasBalancedCleanup(lock)
select lock, "This lock might not be unlocked."
\end{lstlisting}
\vspace{-1.5em}

\textit{\underline{The second style}} suppresses false positives using local semantic evidence around values and control flow. Explicit validation or sanitization checks for validation, normalization, escaping, range checks, and related guards before the reported operation. Implicit program invariant captures value or type constraints that make the reported comparison or arithmetic behavior unreachable in practice. Control-flow dependency captures cases where the reported operation is gated by routing or dispatch logic that prevents the bypass condition from being attacker-controlled in the way assumed by the original query.

\textit{\underline{The third style}} suppresses false positives using project-context evidence. Authorized or privileged actor, trusted value origin, restricted disclosure channel, attacker-unreachable lifecycle, and internal runtime component all depend on recognizing where the reported operation occurs in the application. These predicates use reviewed callables, framework-specific contexts, runtime components, internal service values, logging contexts, or lifecycle methods to encode the evidence used during manual review. Listing~\ref{lst:rq2-auth-refinement} shows a simplified authorized-actor example. It does not infer authorization generally; it suppresses findings only when the reported node is inside a reviewed callable context for a specific rule. This refinement also removes the Kylin \trustedsource{} example in RQ1, where the reported file operation occurs inside an administrator-only workflow.

\begin{lstlisting}[
  language=CodeQL,
  style=codeqlstyle,
  caption={Simplified context-based refinement for authorized or privileged actor false positives.},
  label={lst:rq2-auth-refinement}
]
import java
import semmle.code.java.security.LogInjectionQuery
import LogInjectionFlow::PathGraph
/** Refinement: reviewed privileged contexts. */
predicate isAuthorizedContext(Callable c) {
  c.getDeclaringType().getName() =
      "OAuthProviderProcessingFilter" and
  c.getName() = "validateSignature"
  // Additional reviewed contexts omitted.
}
/** Refinement: node is inside privileged context. */
predicate suppressAuthorizedFinding(DataFlow::Node n) {
  exists(Expr e |
    n.asExpr() = e and
    isAuthorizedContext(e.getEnclosingCallable())
  )
}
from LogInjectionFlow::PathNode source,
     LogInjectionFlow::PathNode sink
where
  LogInjectionFlow::flowPath(source, sink) and
  not suppressAuthorizedFinding(sink.getNode())
select sink.getNode(), source, sink, "..."
\end{lstlisting}

\begin{takeawaybox}
\textbf{RQ2 Summary:} Recurring false-positive patterns can be modeled directly in CodeQL query logic, substantially reducing false positives while preserving most true positives. However, fixed refinements have limited generalization when the pattern depends on project-specific context.
\end{takeawaybox}

\subsection{RQ3: Query Refinement Using LLM Agents}
\label{sec:results-rq3}

Table~\ref{tab:rq3-per-query} reports compile-pass and success results for Codex and Claude Code in guided and unguided settings. A task is successful when the submitted query compiles and removes the target false positive.

Overall, agents struggle to write CodeQL refinements from scratch. In the unguided setting, Codex produces compiling queries for 15/50 (30.0\%) tasks and Claude Code for 18/50 (36.0\%) tasks. Both agents write successful queries in 14/50 (28.0\%) tasks. This suggests that while agents can sometimes produce effective queries, their overall success rate remains low, and many attempts fail due to syntax issues or incomplete query construction that prevent compilation.

Guidance from the RQ2 refinements substantially improves both compilation and success rates. In the guided setting, Codex produces compiling queries for 45/50 (90.0\%) tasks and successful refinements for 28/50 (56.0\%) tasks. Claude Code produces compiling queries for 46/50 (92.0\%) tasks and successful refinements for 31/50 (62.0\%) tasks. Across both agents, guided refinement succeeds at double or more than the unguided success rate. Similarly, the compile-pass rate increases from 33/100 (33.0\%) in the unguided setting to 91/100 (91.0\%) in the guided setting.

\begin{table}[t]
  \centering
  \scriptsize
  \setlength{\tabcolsep}{2pt}
  \renewcommand{\arraystretch}{1.06}
  \caption{RQ3 per-query compile and success results. Each cell reports the query compile (pass) counts.}
  \label{tab:rq3-per-query}
  \begin{tabularx}{\columnwidth}{>{\raggedright\arraybackslash\ttfamily}Xcccc}
  \toprule
  \textbf{Query} &
  \multicolumn{2}{c}{\textbf{Unguided}} &
  \multicolumn{2}{c}{\textbf{Guided}} \\
  \cmidrule(lr){2-3}
  \cmidrule(lr){4-5}
  &
  \textbf{Codex} &
  \textbf{Claude} &
  \textbf{Codex} &
  \textbf{Claude} \\
  \midrule
  \RqThreePerQueryRows
  \midrule
  \RqThreePerQueryTotalRow
  \bottomrule
  \end{tabularx}
  \vspace{-2.7em}
\end{table}

These results show that the RQ2 refinements are useful as adaptation templates. Rather than filtering individual false positives, agents use the refinement logic as a starting point and adjust it to project-specific requirements.
Some guided queries compile but do not remove the target false positive, especially for queries such as \texttt{java/log-injection} and \texttt{java/path-injection}. This suggests that some false positives require project context that is hard for agents to capture and express as a local query adjustment. Regardless, the guided setting consistently improves over unguided refinement for both agents, showing that evidence-based refinement patterns make agent-assisted CodeQL adaptation more effective.

\begin{takeawaybox}
\textbf{RQ3 Summary:} Agentic LLMs are most effective when extending our empirically grounded refinements, not when writing CodeQL refinements from scratch. Our refinements can therefore serve as templates to automatically create project-specific queries for false positive reduction.
\end{takeawaybox}

\section{Related Work}

\noindent
\textbf{Developer experience with SAST false positives. }
Prior work shows that false positives are a key reason why developers distrust or avoid static analysis tools. Developers often ignore warnings that are hard to understand, poorly integrated into workflows, or unactionable~\cite{johnson2013staticanalysis}.
Adoption also depends on practical tool qualities such as good warning phrasing, low false positive rates, workflow integration, suppression support, and relevance to developer priorities~\cite{christakis2016programanalysis}. Code-level suppressions of static analysis warnings are common and often arise from false positives, imprecise warning messages, inaccurate analyzer configuration, third-party code, and developer decisions~\cite{hu2025suppressions}. These studies explain why false positives matter in practice. Our work differs by analyzing why specific CodeQL security paths are false positives, using evidence from source code and trust boundaries, as well as by testing whether recurring causes can guide query refinements.

\noindent
\textbf{Empirical studies of SAST and taint-analysis limitations.}
Several studies analyze static analysis tools and the warnings they produce. Aloraini et al.~\cite{aloraini2019sastwarnings} studied how SAST security warnings differ in their tendency to be false positives. Cui et al.~\cite{cui2024falsepositives} examined false positive and false negative issues in PMD~\cite{pmd2024}, SpotBugs~\cite{spotbugs2024}, and SonarQube~\cite{campbell2013sonarqube} by inspecting issue reports, issue-triggering programs, analyzer specifications, implementations, and fixing patches. Charoenwet et al.~\cite{charoenwet2024securereview} evaluated SAST tools for secure code review using vulnerability contributing commits in C and C++ projects and found that many warnings in vulnerable functions were irrelevant to the reviewed vulnerabilities. Li et al.~\cite{li2024evaluating} evaluated query-based SAST detectability for C/C++ vulnerabilities and examined query-quality issues. Related taint-analysis studies also show that false positives can arise from analyzer abstraction limits, such as infeasible paths, framework modeling, and configuration choices~\cite{luo2019qualitative,arzt2015using,qiu2018analyzing,zhang2021analyzing,mordahl2021impact,wang2020scaling,antoniadis2020static}. These works quantify static analysis limitations across tools, warning types, analyzer issues, or taint-analysis settings. Our work focuses instead on CodeQL security paths in Java CVE projects. Rather than classifying only the reported vulnerability location, we inspect the full reported path, including its source, intermediate flow steps, sink, and surrounding project context. We then organize the explanations into higher-level groups and fine-grained reasons, connecting broad false-positive mechanisms to the concrete evidence observed along each path.

\noindent
\textbf{Validation of SAST alerts.}
Prior work has studied how to audit, validate, rank, or reduce reported static analysis alerts. Svoboda et al.~\cite{svoboda2016alertaudits} proposed a lexicon and rules for reviewing static analysis alerts. Muske and Khedker~\cite{muske2015efficient} used additional static analysis to eliminate false positives by checking whether assertion warnings are valid. Muske and Serebrenik~\cite{muske2022postprocessing} surveyed postprocessing approaches for SAST alarms. Wei et al.~\cite{wei2017oasis} proposed OASIS, which ranks Android Lint warnings by linking warning context to user reviews. Additionally, exploit generation has also been explored as a way of validating potential software vulnerabilities~\cite{axe,nitin2025faultline,ullah2025cve,akhoundali2025eradicating,simsek2025pocgen,huang2026penforge}. Wen et al.~\cite{wen2024automatically} use LLMs to inspect static bug warnings, and Xiong and Zhang~\cite{xiong2026sifting} compare Aider, OpenHands, and SWE-agent for vulnerability false positive filtering on OWASP Benchmark cases and real-world Java projects. QLCoder~\cite{wang2025qlcoder} synthesizes CodeQL queries for security vulnerability detection, while Knighter~\cite{yang2025knighter} uses LLMs to synthesize static-analysis checkers. CVE-Bench~\cite{wang2025cve} and Sec-Bench~\cite{lee2026sec} evaluate LLM agents on real-world software security tasks, including CVE repair and broader security workflows. These works make static analysis and security tooling more useful by supporting manual review, ranking, false positive filtering, and exploitability validation. 
Our work complements these efforts by shifting the focus from alert-level postprocessing and standalone query synthesis to query-level refinement of existing CodeQL rules. We also study whether LLM agents can adapt these refinement patterns to project-specific contexts, producing executable query changes rather than black-box alert verdicts.

\section{Threats to Validity}

\mypara{Internal validity}
False-positive classification depends on reviewer judgment, especially when findings involve project-specific detilas and threat models. We mitigate this by inspecting the full reported path and surrounding project context, recording evidence for each label, and excluding sampled cases that appeared to be plausible true positives. The RQ2 refinements may also overfit to the reviewed findings, so we implement them as CodeQL predicates over recurring program facts rather than file, line, or result-specific suppressions, and evaluate them on the full dataset.

\mypara{External validity}
Our study focuses on CodeQL, Java projects, and the ten queries with the highest false positive rates in our dataset. Thus, our results may not generalize to all languages, tools, query sets, or agents. To mitigate this threat, we chose a practical and representative setting: Java is widely used in server-side software, CodeQL is an industrial SAST framework used in real development workflows, and the selected queries correspond to the highest false positive rates we observed. While our claims are limited to the evaluated setting, the study provides evidence for common SAST use cases and a foundation for future work in broader contexts.

\mypara{Construct validity}
We use benchmark CVE alignment to separate findings that correspond to the known vulnerability from findings that do not align with the benchmark CVE. A non-aligned finding may still be a real vulnerability, so we manually inspected sampled findings before assigning false-positive labels. RQ2 and RQ3 measure true-positive preservation using known benchmark true positives; refinements that preserve these findings could still suppress unknown vulnerabilities outside the benchmark oracle.

\section{Conclusions}

In this paper, we studied whether recurring false positives in Java security SAST can be addressed through CodeQL query refinements. From 500 manually reviewed false positives, we identified five groups of source-level causes and used them to design executable refinements. These refinements removed over 80\% of reviewed false positives and 15.79\% of false positives in the full selected-query dataset while retaining 87.5\% of known true positives. We then evaluated whether agentic coding systems can adapt these refinement patterns to new project-specific false positives. Without guidance, Codex and Claude Code each succeeded on 14/50 tasks. With our refinement patterns, they succeeded on 28/50 and 31/50 tasks, respectively, with compile-pass rates above 90\%. Overall, our results support a SAST workflow in which recurring false positive patterns can be first modeled as query refinements and then adapted to project-specific contexts to reduce repeated manual triage.


\end{document}